\documentclass[RNASS]{aastex62}

\newcommand\TZO{T\.ZO}

\graphicspath{{./}{figures/}}

\shorttitle{HV 2112 is in the SMC}
\shortauthors{McMillan \& Church}

\begin{document}

\title{\textit{Gaia} DR2 confirms that candidate Thorne-\.{Z}ytkow object HV\,2112 is in the SMC}

\correspondingauthor{Paul McMillan}
\email{paul@astro.lu.se}

\author[0000-0002-8861-2620]{Paul J. McMillan}
\affil{Department of Astronomy and Theoretical Physics, Lund Observatory, Box 43, SE-221 00 Lund, Sweden}

\author{Ross P. Church}
\affil{Department of Astronomy and Theoretical Physics, Lund Observatory, Box 43, SE-221 00 Lund, Sweden}

\section{Introduction} \label{sec:intro}

\citet{ThorneZytkow75,ThorneZytkow77} first proposed the existence of super-giant stars containing neutron-star cores, ({\it Thorne-\.Zytkow objects} or \TZO{s}).  In massive \TZO{s} the convective envelope reaches down almost to the neutron star's surface, where extremely hot hydrogen burning can synthesise proton-rich elements in the so-called {\it irp}-process \citep{Cannon93,Podsiadlowski+95}.  However, the expected formation rate of \TZO{s} is  small, and their lifetimes are expected to be very short.  Furthermore, \cite{Fryer+96} suggest that accretion onto the neutron star during \TZO\ formation should convert it into a black hole; the accretion luminosity may also unbind the envelope.  Hence the existence of \TZO{s} remains an open question.

The first \TZO\ candidate is HV\,2112, which was observed by \citet{Levesque14} in a survey of Small Magellanic Cloud (SMC) supergiants.  Unlike the other stars in their survey, its spectrum shows strong lines of rubidium, molybdenum, calcium, and lithium.  \citet{Tout14} discuss two possibilities to synthesise these elements: a \TZO, or a super asymptotic giant branch (SAGB) star.  Both \TZO{s} and SAGB stars could produce molybdenum, rubidium and lithium, but \citet{Tout14} could only explain the calcium enhancement with a \TZO.  A third alternative was proposed by \citet{Maccarone16}, in whose model HV\,2112 is a Milky Way halo binary.  The visible star has previously accreted material enriched in molybdenum and rubidium by {\it s}-processing in the companion's  envelope during the AGB, whilst the calcium comes from natal alpha-enhancement typical for the halo.  They speculate that the lithium could be produced either by thermohaline mixing during accretion, or by nova explosions on the unseen white dwarf companion's surface.

These models imply different distances to HV\,2112. Exotic candidates such as super-AGB stars or \TZO{s} are required to make it luminous enough to be in the SMC.  A first giant branch or AGB star only fits if it is in the Galactic halo.  In neither case could the parallax have been measured hitherto, but the proper motion could be used as a proxy for distance.  \citet{Maccarone16} obtained a proper motion from the Southern Proper Motion Survey \citep{Girard11} which was consistent with a typical halo star, but inconsistent with the mean proper motion of the SMC.  \citet{Worley16} carried out a thorough analysis of several proper motion catalogues and concluded that an SMC location was more likely. 

\section{Data} \label{sec:data}

The recent publication of the astonishingly precise parallaxes and proper motions of $1.3$ billion sources in {\it Gaia}'s data release 2 \citep{GaiaDR2:Summary}  confirms that the proper motion of HV\,2112 is fully consistent with it being a member of the SMC. The kinematics of the SMC were mapped as a scientific demonstration of the {\it Gaia} data by \cite{GaiaDR2:DGGC}.

HV 2112 was found with a simple cone search of the {\it Gaia} archive and has source\_id ``Gaia DR2 4687439865638979712''. The proper motion is $(\mu_\alpha*, \mu_\delta) = (1.116\pm0.072,-1.291\pm0.067)\,\mathrm{mas\,yr}^{-1}$, which is consistent with membership of the SMC. The line-of-sight velocity \citep[measured by][]{Levesque14} is also consistent with SMC membership. The {\it Gaia} DR2 parallax measurement is $-0.202\pm0.045 \,\mathrm{mas}$. The median {\it Gaia} DR2 parallax of stars associated with the SMC is $\sim-0.9\,\mu\mathrm{as}$, but there are spatial variations in parallax on scales of up to $\sim0.1\,\mathrm{mas}$, so this is not far from the expected values when the uncertainty is considered \citep{GaiaDR2:Astrometry, GaiaDR2:DGGC}. The parallax alone appears to rule out any scenario in which HV\,2112 is within a few $\mathrm{kpc}$ of the Sun. 

\begin{figure}
\plottwo{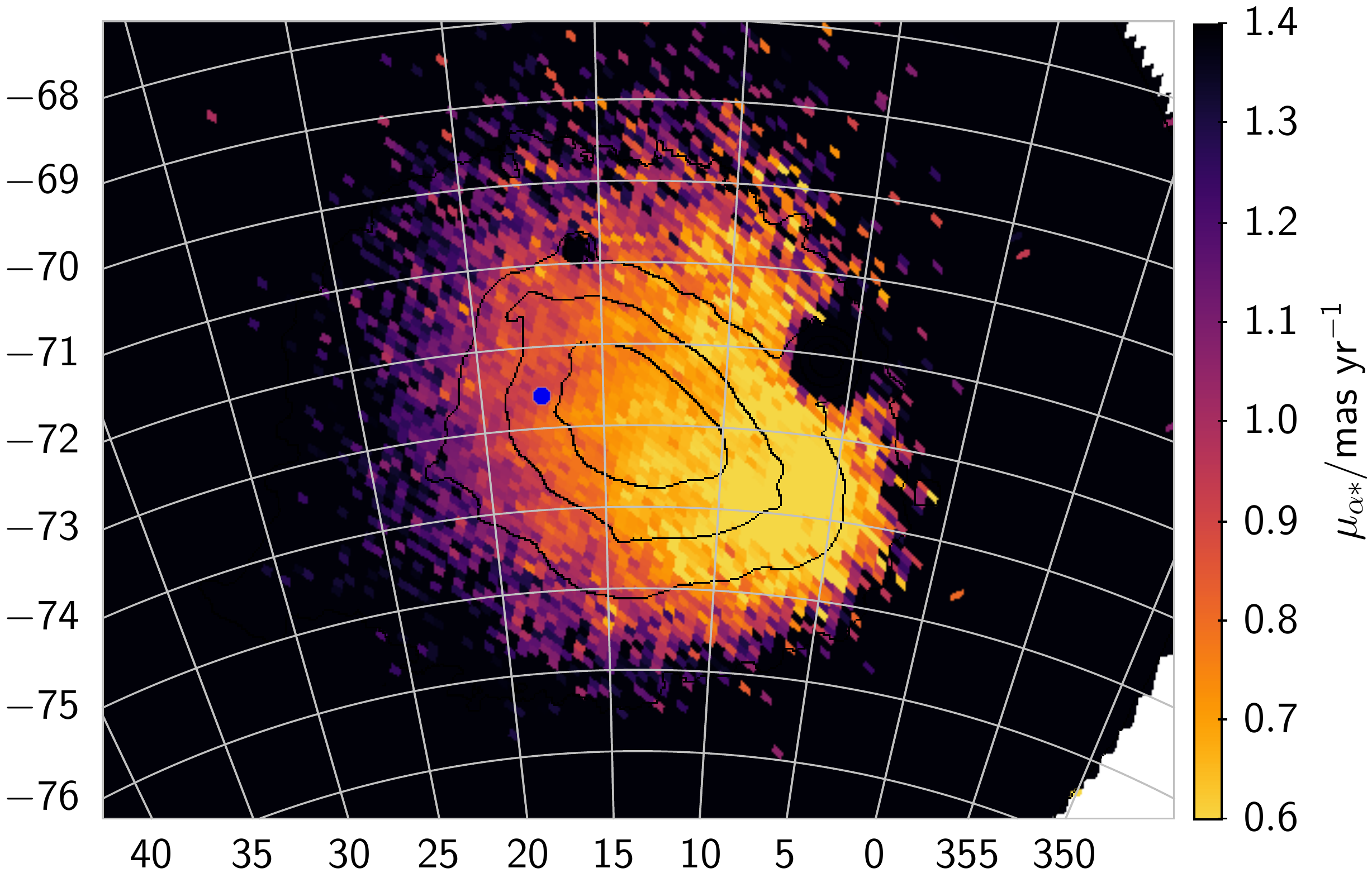}{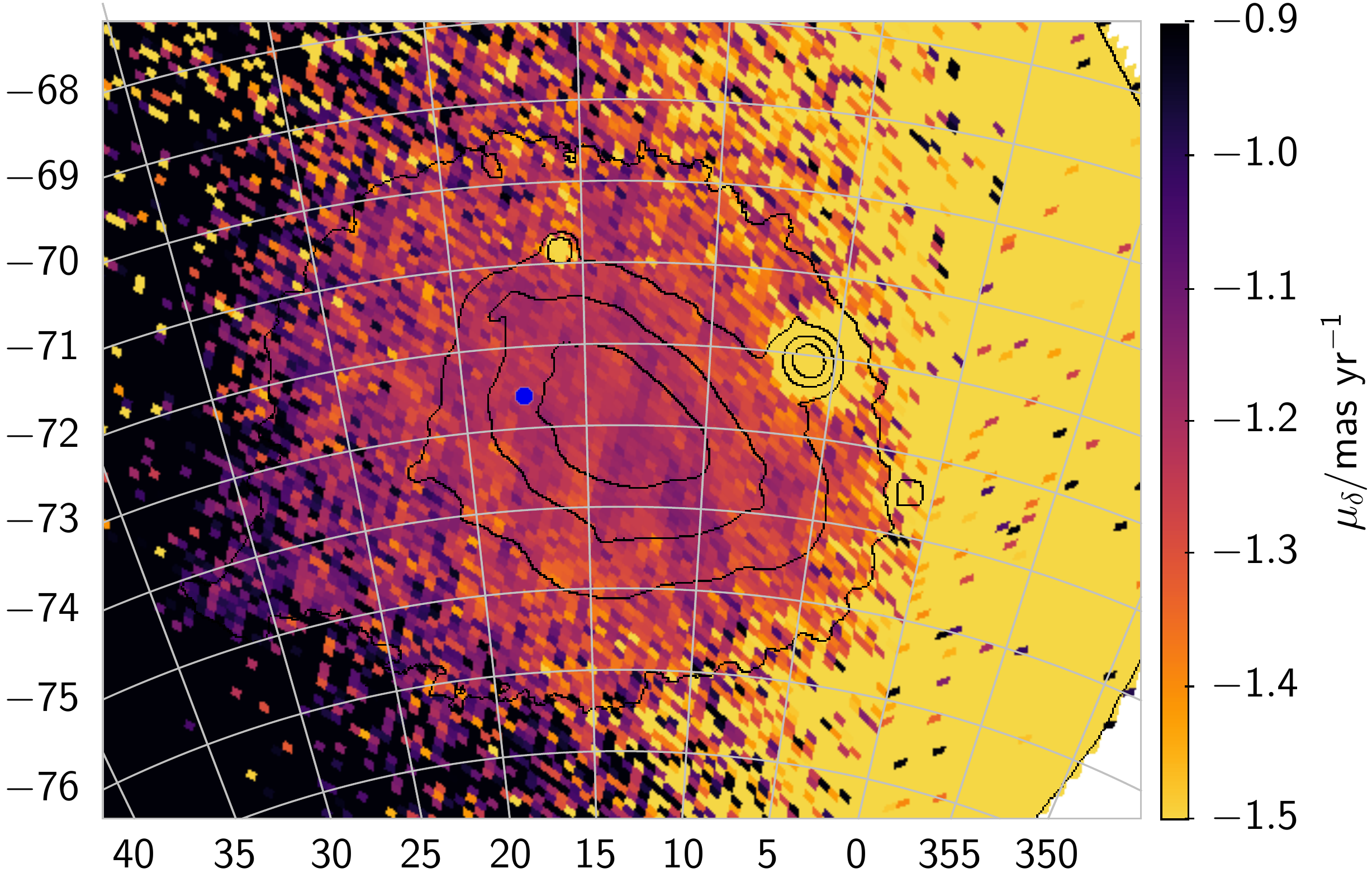}
\caption{The median proper motions $\mu_{\alpha*}$  ({\it left}) and $\mu_\delta$ ({\it right}) as a function of position on the sky (grid labeled with right ascension on the $x$-axis and declination on the $y$-axis) for stars near the SMC. The position of HV 2112 is marked with a blue dot. Proper motions of nearby SMC stars are on average $\approx(1.0,-1.2)\,\mathrm{mas\,yr}^{-1}$, very much consistent with the proper motion of HV 2112. \label{fig:only}}
\end{figure}

In Figure~\ref{fig:only} we show the median proper motions of all {\it Gaia} DR2 sources in the direction of the SMC, with a parallax cut of $\varpi<0.5\,\mathrm{mas}$ to remove foreground contamination. The proper motion of HV\,2112 (blue dot in Figure~\ref{fig:only}) is completely consistent with the proper motion of SMC stars at a similar point in the sky.

We conclude that HV\,2112 is a member of the SMC, and therefore remains the leading candidate \TZO. The 1.3 billion measured parallaxes and proper motions from {\it Gaia} will revolutionise astronomy, and this single measurement already changes our understanding of these exotic objects.

\section*{Acknowledgements}
PJM is grateful to E. Levesque who encouraged him to publish this work. PJM is supported by funding from the Swedish National Space Board. RPC is supported by the Swedish Research Council (grant number 2017-04217).
This work has made use of data from the European Space Agency (ESA)
mission {\it Gaia} (\url{https://www.cosmos.esa.int/gaia}), processed by
the {\it Gaia} Data Processing and Analysis Consortium (DPAC,
\url{https://www.cosmos.esa.int/web/gaia/dpac/consortium}). 

\software{
TOPCAT \citep{TOPCAT}
}

\end{document}